\documentclass[prd,preprintnumbers,twocolumn,eqsecnum,floatfix,letterpaper,superscriptaddress,nofootinbib]{revtex4}
\usepackage{color}
\usepackage{amsmath,amssymb,graphicx}
\usepackage{tensor}
\usepackage{bm}
\usepackage{times}
\usepackage{microtype}
\usepackage{booktabs}
\usepackage{subfigure}
\usepackage[normalem]{ulem}
\usepackage[varg]{txfonts}
\usepackage[colorlinks, pdfborder={0 0 0}]{hyperref}
\definecolor{LinkColor}{rgb}{0.75, 0, 0}
\definecolor{CiteColor}{rgb}{0, 0.5, 0.5}
\definecolor{UrlColor}{rgb}{0, 0, 0.75}
\hypersetup{linkcolor=LinkColor}
\hypersetup{citecolor=CiteColor}
\hypersetup{urlcolor=UrlColor}
\allowdisplaybreaks
\begin{document}

\newcommand{\FigStart}{\begin{figure}[h]}
\newcommand{\FigEnd}{\end{figure}}
\newcommand{\LNh}{\hat{\mathbf{L}}_\text{N}}
\newcommand{\LN}{\mathbf{L}_\text{N}}
\newcommand{\bS}{\mathbf{S}}
\newcommand{\bJ}{\mathbf{J}}
\newcommand{\e}{\mathrm{e}}
\newcommand{\rmi}{\mathrm{i}}
\newcommand{\flow}{f_0}
\newcommand{\fcut}{f_\mathrm{cut}}
\newcommand{\bchi}{\bm{\chi}}
\newcommand{\blambda}{\bm{\lambda}}
\newcommand{\bLambda}{\bm{\Lambda}}
\newcommand{\bchia}{\bm{\chi}_\text{a}}
\newcommand{\bchis}{\bm{\chi}_\text{s}}
\newcommand{\chis}{\chi_\text{s}}
\newcommand{\chia}{\chi_\text{a}}
\newcommand{\chiadL}{\bchia \cdot \LNh}
\newcommand{\chisdL}{\bchis \cdot \LNh}
\newcommand{\chisSqr}{\bchis^2}
\newcommand{\chiaSqr}{\bchia^2}
\newcommand{\chisDchia}{\bchis \cdot \bchia}
\newcommand{\cA}{\mathcal{A}}
\newcommand{\cB}{\mathcal{B}}
\newcommand{\cC}{\mathcal{C}}
\newcommand{\cP}{\mathcal{P}}
\newcommand{\Mc}{M_\mathrm{c}}
\newcommand{\thetaz}{\theta_{0}}
\newcommand{\thetat}{\theta_{3}}
\newcommand{\thetats}{\theta_\mathrm{3S}}
\newcommand{\psiL}{\psi^\text{L}}
\newcommand{\dphi}{\partial \psi}
\newcommand{\dtheta}{\partial \theta}
\newcommand{\dphiL}{\partial \psi^\text{L}}
\newcommand{\derb}{\partial_b}
\newcommand{\dera}{\partial_a}
\newcommand{\df}{{\mathrm{d}f}}
\newcommand{\match}{\mathcal{M}}
\newcommand{\bOmega}{\mathbf{\Omega}}
\newcommand{\btheta}{\bm{\theta}}
\newcommand{\SBank}{\textsc{SBank}}
\newcommand{\LALSuite}{\textsc{LALSuite}}
\newcommand{\FFe}{\mathrm{FF_{eff}}}
\newcommand{\FF}{\mathrm{FF}}
\newcommand{\bg}{\mathbf{g}}
\newcommand{\Mchirp}{\mathcal{M}_\text{c}}
\newcommand{\MM}{\mathcal{M}}
\newcommand{\blue}{\color{blue}}
\newcommand{\red}{\color{red}}
\newcommand{\imrphenb}{\textsc{IMRPhenomB}}	
\newcommand{\eobnr}{\textsc{EOBNRv2}} 
\newcommand{\SqrtMet}{$\sqrt{|\textbf{g}|}$ } 
\newcommand{\msun}{$M_{\odot}$} 

\newcommand{\CMI}{Chennai Mathematical Institute, Plot H1, SIPCOT IT Park, Siruseri, 603103 Tamilnadu, India}
\newcommand{\ICTS}{International Centre for Theoretical Sciences, Tata Institute of Fundamental Research, Bangalore 560012, India}

\title{Template-space metric for searches for gravitational waves\\from the inspiral, merger, and ringdown of binary black holes}

\author{Chinmay Kalaghatgi}
\affiliation{\CMI}
\author{Parameswaran Ajith}
\affiliation{\ICTS}
\author{K. G. Arun}
\affiliation{\CMI}
\begin{abstract}
Searches for gravitational waves (GWs) from binary black holes using interferometric GW detectors require the construction of template banks for performing matched filtering while analyzing the data. Placement of templates over the parameter space of binaries, as well as coincidence tests of GW triggers from multiple detectors make use of the definition of a metric over the space of gravitational waveforms. Although recent searches have employed waveform templates coherently describing the inspiral, merger and ringdown (IMR) of the coalescence, the metric used in the template banks and coincidence tests was derived from post-Newtonian inspiral waveforms. In this paper, we compute (semianalytically) the template-space metric of the IMR waveform family \imrphenb\ over the parameter space of masses and the effective spin parameter. We also propose a coordinate system, which is a modified version of post-Newtonian chirp time coordinates, in which the metric is slowly varying over the parameter space. The match function semianalytically computed using the metric has excellent agreement with the ``exact'' match function computed numerically. We show that the metric is able to provide a reasonable approximation to the match function of other IMR waveform families, such that the effective-one-body model calibrated to numerical relativity (\eobnr). The availability of this metric can contribute to improving the sensitivity of searches for GWs from binary black holes in the advanced detector era. 
\end{abstract}

\preprint{LIGO-P1400252-v3}
\maketitle

\section{Introduction}
Second-generation laser interferometric gravitational-wave (GW) detectors such as Advanced LIGO~\cite{0264-9381-27-8-084006} and Advanced Virgo~\cite{AdvVirgo:2009} are expected to start their first data-taking runs by 2015 and are expected to reach their design sensitivities in a few years (see~\cite{Aasi:2013wya} for a discussion of likely observing scenarios). In addition, an advanced interferometric detector, called KAGRA~\cite{Somiya-Kagra:2012}, is being constructed in Japan, and another one in India, called LIGO-India~\cite{LIGOIndiaProposal:2011} is expected to be built in India. Coalescence of binary black holes (BBHs) are among the most promising sources for the first direct detection of GWs. With the designed sensitivities of these detectors, anticipated detection rate of BBH coalescences is $\sim 0.4 - 1000$ per year~\cite{2010CQGra..27q3001A}. Observation of GWs from BBHs is expected to make significant contributions to our understanding of fundamental physics, astrophysics and cosmology (see \cite{SathyaSchutzLivRev09} for a review). 

The expected gravitational waveforms from BBHs can be accurately computed from General Relativity using appropriate approximation techniques or numerical methods (see, e.g.~\cite{Blanchet:LivRev,Centrella:2010mx,Berti:2009kk}, for reviews). Different ``template'' waveforms corresponding to different parameters of the binary (such as the component masses and spin angular momenta of the black holes), are cross-correlated with the data looking for correlations that exceed certain threshold, indicating the presence of a GW signal. This technique is called \emph{matched filtering}. Additional signal consistency tests and multi-detector coincidence tests are employed to further assess the true nature of the signal (see~\cite{Babak:2012zx} for a detailed discussion). Several searches for BBHs have been performed in the past using data from the previous science runs of LIGO and Virgo~\cite{Aasi:2012rja,Abadie:2011kd}. The non-detection by these initial instruments is consistent with our expectation of the astrophysical rates of BBH coalescences~\cite{2010CQGra..27q3001A}. 

Coalescence of BBHs typically involve three stages: In the early \emph{inspiral} stage, the radial velocity of the black holes is much smaller than their tangential velocity, which itself is much smaller than the velocity of light ($v_r \ll v_\varphi \ll c$). Gravitational radiation reaction causes the binary orbit to continuously shrink and eventually the black holes move with relativistic velocities and ultimately \emph{merge} with each other. In the final \emph{ring down} stage, the merger remnant settles into a Kerr black hole by radiating a spectrum of quasi-normal modes. Accurate analytical models of expected GW signals from the inspiral (see \cite{Bliving} for a review) and ring-down (see \cite{TSLivRev03} for a review) stages are available for the last two decades. The analytical approximation methods cannot be applied to the merger stage, where the gravity is strong and highly non-linear. Due to this, the first searches for GWs from BBHs were performed employing templates either describing the inspiral stage~\cite{Abbott:2007xi,Abbott:2007ai} or the ring-down stage~\cite{Abbott:2009km,Aasi:2014bqj}. In the case of BBHs consisting of intermediate-mass ($m_1, m_2 \gtrsim 100 M_\odot$), only the merger and ring-down stages of the coalescence will be observable in ground-based detectors. Such signals appear as bursts of gravitational radiation. Several searches have been performed in the past that look for short-lived excess power in the data~\cite{VirgoC7,Virgo:2012aa,Aasi:2014iwa}. 

Recent breakthroughs in numerical relativity~\cite{Pretorius:2005gq,Campanelli:2005dd,Baker05a} have enabled us to accurately compute the expected gravitational waveforms from the merger stage. Large catalogs of numerical-relativity waveforms have become available now~\cite{Ajith:2012az,SXS-Catalog,Mroue:2013xna}. This has lead to the development of several analytical models coherently describing the GW signals from the inspiral, merger and ring-down (IMR)~\cite{Buonanno:2007pf,Pan:2011gk,Taracchini:2013rva,Pan:2013rra,Taracchini:2012ig,Damour:2007vq,Damour:2008te,Damour:2009kr,Damour:2012ky,Damour:2014sva,Ajith:2007qp,Ajith:2009bn,Ajith:2007xh,Ajith:2007kx,Santamaria:2010yb,Hannam:2013oca}. Recent LIGO-Virgo searches for GWs from BBHs made use of these IMR templates~\cite{Aasi:2012rja,Abadie:2011kd}, which significantly improved the sensitivity of these searches. 

Since the parameters of potential the GW signals buried in the detector data are not known \emph{a priori}, the data has to be cross-correlated with a ``bank'' of template waveforms corresponding to different parameters. A discrete set of template parameters has to be chosen such that, for any signal, there is always a template that is sufficiently close to it. At the same time, in order to minimize the computational burden of the search, it is desirable to keep the number of templates in the bank to a minimum. In order to address this \emph{covering problem} in template placement, a geometrical method has been developed~\cite{Sathyaprakash:1991mt,Owen:1995tm,Owen:1998dk}. In this method, templates are placed in the parameter space such that inner product between neighboring templates is fixed to a predetermined value (called the \emph{minimal match}~\cite{Owen:1995tm}), say 0.97. This ensures that the loss of signal-to-noise ratio (SNR) due to the mismatch between the signal and the closest template in the bank is acceptably small, say 3\%. 

This geometric formalism introduces the notion of a metric in the space of GW signals, which allows us to place templates in the parameter space employing lattice-based methods~\cite{hexabank,Babak:2006ty,Brown:2012qf,Harry:2013tca} or stochastic placement methods~\cite{Babak:2008rb,Harry:2009,2010PhRvD81b4004M,Ajith:2012mn}. In addition, the multi-detector coincidence test~\cite{Robinson:2008un} employed in these searches also requires the knowledge of the metric. The metric can be computed from the template waveforms using the formalism introduced in~\cite{Owen:1995tm}. Metrics in the space of post-Newtonian (PN) waveforms describing the inspiral part of the coalescence have been computed in the past~\cite{Owen:1995tm,Tanaka:2000xy,Keppel:2013kia,Ajith:2012mn}. In the past searches for GWs from BBHs, even those employing IMR templates, the PN metric was employed in the construction of template banks and in the multi-detector coincidence tests. Although the degradation of the SNR due to this choice of the metric was not drastic, future searches will greatly benefit from the knowledge of the actual metric in the space of IMR waveforms. 

In this paper, we compute the metric in the space of the IMR waveform family \imrphenb~\cite{Ajith:2009bn}. We identify a coordinate system in which the metric is slowly varying over the parameter space of interest, which is desirable in the construction of template banks. We show that the inner product between the waveforms (known as the \emph{match}) computed using the metric has excellent agreement with the exact numerical computation of the match. We also show that the match function computed using the \imrphenb\ metric agrees well with that computed numerically from other IMR waveform families, such as the \eobnr~\cite{Pan:2011gk}. This metric can be employed in the construction of template banks as well as in multi-detector coincidence tests. We expect that this will significantly contribute to improving the sensitivity of searches of GWs from BBHs in the advanced detector era. 

The rest of the paper is organized as follows: Section~\ref{sec:geom} provides a brief overview of the geometrical approach employed in the template placement, introducing the notion of the metric. Section~\ref{sec:metric_calc} describes the calculation of the metric for the \imrphenb\ family and discusses the coordinates in terms of which the metric is more or less uniform. A discussion of the results, in particular the comparison of the match function computed using the metric with exact match function, is provided in Sec.~\ref{sec:results}, while Sec.~\ref{sec:conclusions} provides some concluding remarks and lists the future directions. 

\section{Geometrical approach to template placement}\label{sec:geom}

Here we provide a brief overview of the metric formalism originally introduced in~\cite{Owen:1995tm} for laying down waveform templates in the parameter space of compact binaries. A set of intrinsic and extrinsic parameters $\boldsymbol{\lambda} = \{\boldsymbol{\lambda}_{\mathrm{extrinsic}}, \boldsymbol{\lambda}_{\mathrm{intrinsic}}\}$ parameterize a gravitational waveform $h(f;\boldsymbol{\lambda})$. The intrinsic parameters are parameters that are intrinsic to the source, such as the masses and spins of the compact objects, while the extrinsic parameters are those which depend on the relative location of the source with respect to the detector (such as the time of arrival $t_0$ of the signal at the detector and the phase of the signal $\varphi_0$ at a reference time $t_0$~\footnote{In the case of binaries not exhibiting spin precession, if we neglect the effect of higher harmonics (no-quadrupole modes), it can be shown that all other extrinsic parameters are degenerate with the parameters $t_0$ and $\varphi_0$, and can be absorbed into these.}.

The \textit{match} between any two waveforms ${h}(f, \boldsymbol{\lambda})$ and ${h}(f, \boldsymbol{\lambda} + \Delta \boldsymbol{\lambda})$ is defined as:
\begin{equation}
 \mathcal{M}(\boldsymbol{\lambda}, \boldsymbol{\lambda} + \Delta \boldsymbol{\lambda}) \equiv \mathrm{max}_{\Delta \boldsymbol{\lambda}_{\mathrm{extrinsic}}} \left< \hat{h}(f,\boldsymbol{\lambda}) , \hat{h}(f, \boldsymbol{\lambda} + \Delta \boldsymbol{\lambda}) \right>,
\end{equation}
where $\left< a, b \right>$ denote the noise weighted inner product: 
\begin{equation}
 \left< a, b \right> \equiv 2 \int_{f_{0}}^{\infty} \frac{a(f)b^{*}(f) + b(f)a^{*}(f)}{S_{h}(f)}
\end{equation}
 where $S_{h}(f)$ is the one-sided power spectral density of the detector noise, $f_0$ is the low-frequency cutoff of the detector, and a ``hat'' denotes a normalized waveform: $\hat{h} \equiv  h/||h||$ where $||h|| \equiv \sqrt{\left<h, h\right>}$. 

The match function has its maximum value ($ \mathcal{M}_{max}$ = 1) at $\Delta \boldsymbol{\lambda} = 0$. Taylor-expanding the match function about $\Delta \boldsymbol{\lambda} = 0$ up to quadratic order gives: 
\begin{equation}  \label{eq:MatchDef}
 \mathcal{M}(\boldsymbol{\lambda}, \boldsymbol{\lambda} + \Delta \boldsymbol{\lambda}) \simeq 1  + \frac{1}{2} \left( \frac{\partial^{2} \mathcal{M}}{\partial \Delta \lambda^{i} \partial \Delta \lambda^{j}} \right) \Delta \lambda^{i} \Delta \lambda^{j}.
\end{equation}

The \emph{mismatch} $1 -  \mathcal{M}(\boldsymbol{\lambda}, \boldsymbol{\lambda} + \Delta \boldsymbol{\lambda})$ can be thought of as the proper distance between points $\boldsymbol{\lambda}$ and $\boldsymbol{\lambda} + \Delta \boldsymbol{\lambda}$ (in the signal manifold), and can be written as 
\begin{equation}\label{eq:MismatchQuad}
1 - \mathcal{M} \simeq g_{ij} \, \Delta \lambda^{i} \Delta \lambda^{j}. 
\end{equation}
This introduces the notion of a metric in the parameter space, defined as 
\begin{equation}
 g_{ij} \equiv - \frac{1}{2} \left( \frac{\partial^{2} \mathcal{M}}{\partial \Delta \lambda^{i} \partial \Delta \lambda^{j}} \right)
\end{equation}

The metric over the intrinsic parameter space can be calculated from the \emph{Fisher information matrix} by projecting it on to the subspace orthogonal to the space of extrinsic parameters. The Fisher information matrix is defined as:
\begin{equation}
 \Gamma_{ij} = \frac{1}{2} \, \left< \partial_{i} \hat{h}(f; \boldsymbol{\lambda}), \partial_{j} \hat{h}(f; \boldsymbol{\lambda}) \right>, \label{eq:Fisher}
\end{equation}
where $\partial_{i}$ denotes partial derivative w.r.t. the parameter $\lambda_{i}$. The metric over the three-dimensional space of intrinsic parameters can be computed from this as 
\begin{equation}
 \textbf{g} = \boldsymbol{\Gamma}_{1} - \boldsymbol{\Gamma}_{2}^\mathrm{T} \, \boldsymbol{\Gamma}_{3}^{-1} \, \boldsymbol{\Gamma}_{2},\label{eq:metricproj}
\end{equation}
where $\boldsymbol{\Gamma}_{1}$ is the Fisher matrix over the intrinsic parameters, $\boldsymbol{\Gamma}_{3}$ the same over the extrinsic parameters, and $\boldsymbol{\Gamma}_{2}$ that describe the cross terms. Assuming three intrinsic parameters (say, total mass, mass ratio and one spin parameter as described in Sec~\ref{sec:PhenomB}), and two extrinsic parameters (reference time and phase), $\boldsymbol{\Gamma}_{1}, \boldsymbol{\Gamma}_{2}, \boldsymbol{\Gamma}_{3}$  are defined, respectively, as 
\begin{eqnarray}
\left[
\begin{array}{ccc}
\Gamma_{11} & \Gamma_{12} & \Gamma_{13} \\
\Gamma_{21} & \Gamma_{22} & \Gamma_{23} \\
\Gamma_{31} & \Gamma_{32} & \Gamma_{33}
\end{array}
\right],
\left[
\begin{array}{ccc}
\Gamma_{41} & \Gamma_{42} & \Gamma_{43}  \\
\Gamma_{51} & \Gamma_{52} & \Gamma_{53}  
\end{array}
\right], 
\left[
\begin{array}{cc}
\Gamma_{44} & \Gamma_{45}  \\
\Gamma_{54} & \Gamma_{55}  \\
\end{array}
\right].
\end{eqnarray}
Similarly for a two dimensional metric corresponding to, say, the two mass parameters,  $\boldsymbol{\Gamma}_{1}, \boldsymbol{\Gamma}_{2}, \boldsymbol{\Gamma}_{3}$ read
\begin{eqnarray}
\left[
\begin{array}{cc}
\Gamma_{11} & \Gamma_{12}  \\
\Gamma_{21} & \Gamma_{22}  
\end{array}
\right],
\left[
\begin{array}{cc}
\Gamma_{31} & \Gamma_{32}  \\
\Gamma_{41} & \Gamma_{42} \\ 
\end{array}
\right], 
\left[
\begin{array}{cc}
\Gamma_{33} & \Gamma_{34}  \\
\Gamma_{43} & \Gamma_{44}  \\
\end{array}
\right],
\end{eqnarray}
respectively.
\section{Computation of the metric in the space of the inspiral, merger, ringdown waveforms}
\label{sec:metric_calc}

\subsection{IMRPhenomB waveform}\label{sec:PhenomB}
The \imrphenb\ waveform models GW signals from the inspiral, merger and ringdown phases of a coalescing black hole binary with non-precessing 
spins~\cite{Ajith:2009bn}. The waveform is described by three intrinsic parameters: the total mass $M \equiv m_1 + m_2$, the symmetric mass ratio $\eta \equiv m_1 m_2 / M^2$, and a single effective spin parameter $\chi \equiv (m_1 \, \chi_1 + m_2 \, \chi_2)/M$, where $\chi_{1,2}$ are the dimensionless spins of the two black holes and $m_{1,2}$ are the masses of the binary components. The waveform is written as $h(f) \equiv A(f) \, e^{-\mathrm{i} \Psi(f)}$ where the amplitude $A(f)$ is  defined as 
\begin{equation} \label{eq:IMRAmp}
 A(f) \equiv \mathcal{C}\,f_1^{-7/6}  
\begin{cases}
 f'^{-7/6} \left[ 1 + \sum_{i=2}^{3} \, \alpha_{i} \, v^{i} \right] &  f_0 \leq f < f_{1} \\
                                  w_{m} f'^{-2/3} \left[ 1 + \sum_{i=1}^{2} \, \epsilon_{i} \, v^{i} \right] & f_{1} \leq f < f_{2}\\
                                  w_{r} \mathcal{L} (f,f_{2},\sigma) & f_{2} \leq f < f_{3} 
 \end{cases}
\end{equation}
where $\mathcal{C} = \frac{M^{5/6}}{d \pi^{2/3}} \left( \frac{5 \eta}{24}\right) ^{1/2}$ and $d$ is the \emph{effective distance} to the source~\footnote{Effective distance is a combination of the luminosity distance, the antenna pattern functions of the detector and the inclination angle of the binary, which determines the observed amplitude of GW signals in the detector.}. The frequencies $f_{1}, f_{2}$ are the transition frequencies between the inspiral, merger and merger, ringdown stages. The amplitude is zero below the cutoff frequency $f_0$ and above the cutoff frequency $f_3$. The other quantities appearing in the above expression are,  $f' \equiv f/f_{1}$, $v \equiv (\pi M f)^{1/3}$, $\mathcal{L}(f,f_2, \sigma)$ is a Lorentzian function with width $\sigma$ centered around the frequency $f_{2}$. The normalization constants $w_{m}$ and $w_{r}$ make the amplitude smooth over the transition frequencies $f_{1}$ and $f_{2}$. The phenomenologically introduced parameters $\epsilon_{1} = 1.4547\chi - 1.8897$ and $\epsilon_{2} = -1.8153\chi + 1.6557$ model the amplitude of the merger part, while $\alpha_{2} = -323/224 + 451\eta /168$ and $\alpha_{3} = (27/8 - 11\eta/6)$ are the 1.5PN accurate post-Newtonian corrections to the inspiral amplitude. 

The phase of the waveform is given by
\begin{equation} \label{eq:IMRPhase}
 \Psi(f) \equiv 2\pi ft_{0} + \phi_{0} + \frac{3}{128 \, \eta \, v^{5}}\left[ 1 + \sum_{k=2}^{7} \, \psi_k \, v^{k} \right], 
\end{equation}
where $t_{0}$ and $\phi_{0}$ are the time of arrival of the signal and the corresponding phase
by $t_{0}$ and $\phi_{0}$, respectively. The phenomenologically calibrated parameters $\psi_k$ describes the phase evolution of the binary and are given in Table I of \cite{Ajith:2009bn}. 

\subsection{Choice of coordinates}\label{sec:NewCord}

\begin{figure*}[ht]\includegraphics[width = 0.89\textwidth]{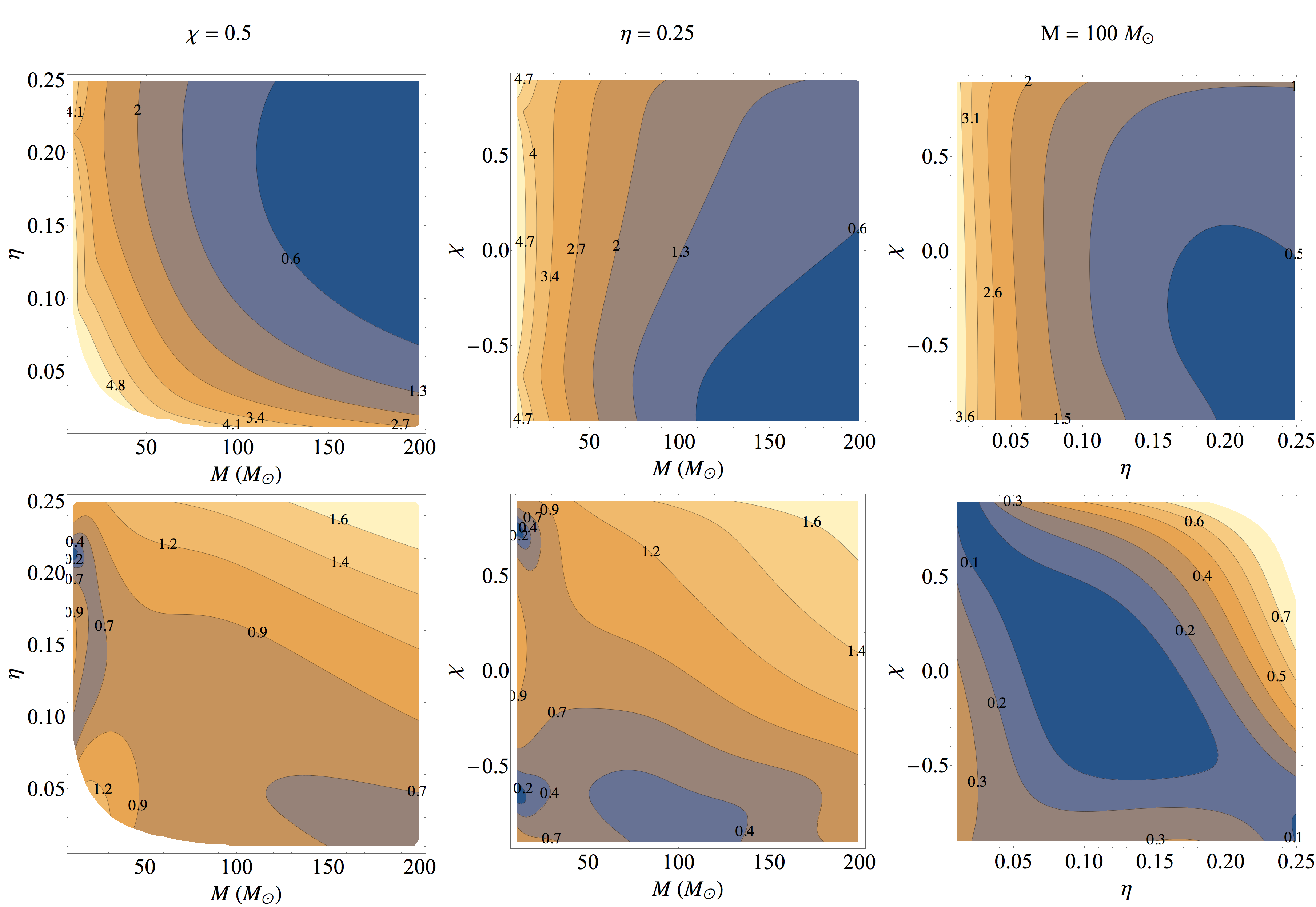}\caption{Variation of the square root of the determinant of the \imrphenb\ metric over the parameter space. The plots show the contours of $\frac{1}{2} (\log |\textbf{g}| - \log |\textbf{g}|_\mathrm{min})$, where $|\textbf{g}|_\mathrm{min}$  is the minimum value of $|\textbf{g}|$ in each panel. In the top panel, the metric  is computed in the $\{M, \eta, \chi\}$ coordinate system, while in the bottom panel the metric is computed in the $\{\xi_{0}, \xi_{3}, \xi_{3S}\}$ coordinate system [see Eq.~(\ref{eq:xi}) for definition]. It can be seen that the variation of the quantity over the parameter space in the bottom panels is significantly smaller than that in the top panels.}\label{fig:MetricVariation}
\end{figure*}

It is desirable to have a coordinate system in which the metric is (at least nearly) constant over the parameter space. For example, the approximation for the mismatch given in Eq.~(\ref{eq:MismatchQuad}) assumes that the metric is nearly constant over the two points in the parameter space for which the match is computed, and hence is slowly varying over the parameter space. Additionally, knowledge of a coordinate system in which the local density of templates (which is proportional $\sqrt{|\mathbf{g}|}$) over the parameter space is nearly constant is greatly helpful for stochastic placement methods, since the new random ``proposals'' of the templates can be drawn from a simple uniform distribution~\cite{Ajith:2012mn}. The conventional ``physical'' parameters of the binary, such as $\{M, \eta, \chi\}$ do not form a nice coordinate system for this purpose, since $\sqrt{|\mathbf{g}|}$ varies by at least 4 orders of magnitude over the parameter space of interest ($1 M_\odot \lesssim m_1, m_2$, $m_1 + m_2 \gtrsim 10 M_\odot$, $-0.95 \lesssim \chi \lesssim 0.95$). A set of coordinates termed \emph{chirp times} were introduced by~\cite{Sathyaprakash:1991mt} in which the non-spinning PN metric is slowly varying. This was generalized to the case of non-precessing spins by~\cite{Ajith:2012mn}. The square root of the determinant of the PN metric computed in this coordinate system is found to have variation $\lesssim 30$ over the ``low-mass'' region in the parameter space ($1 M_\odot \lesssim m_1,m_2 \lesssim 20 M_\odot$)~\cite{Ajith:2012mn}. 

We have found that in the PN chirp time coordinate system, the square root of the determinant of the \imrphenb\ metric has a variation of $\sim 100$ over the parameter space of interest  ($1 M_\odot \lesssim m_1, m_2$, $m_1 + m_2 \gtrsim 10 M_\odot$, $-0.95 \lesssim \chi \lesssim 0.95$). Here we introduce a new coordinate system $\{\xi_{0}, \xi_{3}, \xi_{3S}\}$, which is a modified version of the PN chirp time coordinates. 
\begin{eqnarray}\label{eq:xi}
\xi_{0} &\equiv& \frac{5}{128 \, \eta \, (\pi M f_{0})^{5/3}}, \\ 
\xi_{3} &\equiv&  \frac{\pi}{4 \, \eta \, (\pi M f_{0})^{2/3}}, \\
\xi_{3S} &\equiv&  \frac{\pi \,  (17022 - 9565.9\, \chi )}{4 \, (\pi M f_{0})^{2/3}}.
\end{eqnarray}
It can be seen that $\xi_{0}$ and $\xi_{3}$ are nothing but the dimensionless chirp times introduced in~\cite{Owen:1998dk}: 
\begin{equation}
\xi_{0} = 2 \pi f_0 \tau_0, ~~~ \xi_{3} = -2 \pi f_0 \tau_3,
\end{equation}
where $\tau_0$ and $\tau_3$ are the familiar Newtonian and 1.5PN chirp times~\cite{Sathyaprakash:1991mt,PhysRevD.57.630}, and $f_0$ is the low-frequency cutoff of the detector noise.

The PN chirp time coordinates are judiciously chosen to make the corresponding PN coefficients in the phasing formula (at 0PN and 1.5PN order) linear in these coordinates. For the case of nonspinning \imrphenb\ waveforms, the 0PN term in the phase is the same as that of the PN waveform. Also the $\eta-$ independent coefficient (test-mass limit) in the 1.5PN term is the same as that of the PN waveform (see Table 1 of \cite{Ajith:2009bn}). Hence the corresponding (dimensionless) PN chirp times ($\xi_0,\xi_3$) make the metric nearly constant over the two dimensional space of mass parameters. But, when spins are included, the 1.5PN coefficient of the \imrphenb\ phase has terms up to quadratic order in $\eta$ and $\chi$. Further, for ease of computations, the coordinates have to be invertible and the transformation has to be bijective. With these aims, we write $\xi_{3S}$ as a product of $\xi_{3}$ and the terms linear in $\eta$ and $\eta\,\chi$ in the 1.5PN order coefficient of \imrphenb\ waveform phase. Admittedly, our procedure of finding the new coordinate system is somewhat ad-hoc and there may be a better coordinate system in which the variation of the metric is even smaller. However, we believe that this coordinate system is adequate for the purposes of template placement and coincidence tests.

The physical parameters can then be written in terms of our new coordinates $\{\xi_{0}, \xi_{3}, \xi_{3S}\}$ as 
\begin{eqnarray}
M &=& \frac{5}{32 \pi^{2} f_{0}} \frac{\xi_{3}}{\xi_{0}} \\
\eta &=& \left( \frac{16 \pi^{5}}{25} \right)^{1/3} \frac{\xi_{0}^{2/3}}{\xi_{3}^{5/3}} \\
\chi &=& \frac{17022}{9565.9} - \frac{\xi_{3S}}{9565.9} \left( \frac{25 \, \xi_3^2}{16 \, \pi^{5} \, \xi_0^2} \right)^{1/3} 
\end{eqnarray}
In this coordinate system, \SqrtMet  has a maximum variation of $\lesssim 40$  over the parameter space of interest. Figure~\ref{fig:MetricVariation} provides a comparison of the \SqrtMet as computed in the two different coordinate systems. 

\subsection{Computation of the metric}

The codes to compute the metric are available as part of the \textsc{LALSimulation} package, which is part of the LSC Algorithms Library~\cite{LAL} -- the core software package used for GW data analysis by the LIGO-Virgo collaborations. Here we provide some details of the numerical implementation of the computation of the metric in \textsc{LALSimulation}. The \imrphenb\ waveform amplitude described in Sec.~(\ref{sec:PhenomB}) can be rewritten as 
\begin{equation} \label{eq:IMRAmp2}
 A(f) \equiv  
\begin{cases}
A_1(f)\,, &  f_0 \leq f < f_{1} \\
A_2(f)\,, & f_{1} \leq f < f_{2}\\
A_3(f)\,, & f_{2} \leq f < f_{3}\,.
\end{cases}
\end{equation}
Above, 
\begin{eqnarray}
A_1(f) & = & \sum_{k=0}^{3} \, \beta_k (\blambda) \, f^{(2k-7)/6}; ~~~ \beta_k \equiv \mathcal{C} \alpha_k \, (\pi M)^{k/3}, \nonumber \\ 
A_2(f) & = & \sum_{k=0}^{2} \, \kappa_k (\blambda) \, f^{(k-2)/3}; ~~~ \kappa_k \equiv \mathcal{C} w_m  \epsilon_k \, (\pi M)^{k/3} f_1^{-1/2}, \nonumber \\ 
A_3(f) & = & \xi(\blambda) \, \mathcal{L}(f,f_2,\sigma); ~~~ \xi\equiv \mathcal{C} w_r f_1^{-7/6}\,
\label{eq:phen_ampl_2}
\end{eqnarray}
where $\blambda \equiv \{M, \eta, \chi\}$ is the set of intrinsic parameters~\footnote{Note that $\alpha_1 = 0$ in Eq.~(\ref{eq:phen_ampl_2}).}. Similarly, the phase can be written as 
\begin{equation}
\Psi(f) = \sum_{k=0}^7 \, \varphi_k(\blambda)\, f^{(k-5)/3}\,, 
\end{equation}
where 
\begin{eqnarray}
\varphi_k & = & \frac{3}{128\eta} \psi_k \, (\pi M)^{(k-5)/3}~~~ \mathrm{for}~ k \neq 5,  \nonumber \\
\varphi_5 & = & \frac{128 \, \eta} {3}.
\end{eqnarray}
This allows us to rewrite the Fisher matrix defined in Eq.~(\ref{eq:Fisher}) as 
\begin{equation}
\label{eq:metric_v2}
\Gamma_{ij} \simeq \frac{1}{2 \, ||h||^2} \, \sum_{a = 1}^3 \Big[ \left< A_a \, \partial_i \Psi,  A_a \, \partial_j \Psi \right> +  \left< \partial_i A_a,  \partial_j A_a \right> \Big]\,,
\end{equation}
where 
\begin{eqnarray}
\partial_i A_1(f) & = & \sum_{k=0}^3 \, \partial_i \, \beta_k(\blambda) \, f^{(2k-7)/6} \,, \nonumber \\
\partial_i A_2(f) & = & \sum_{k=0}^2 \, \partial_i \, \kappa_k(\blambda) \, f^{(k-2)/3} \,, \nonumber \\
\partial_i A_3(f) & = & \xi(\blambda) \, \partial_i \, \mathcal{L}(\blambda, f) + \partial_i \, \xi(\blambda) \, \mathcal{L}(\blambda, f)  \,, \nonumber \\
\partial_i \Psi(f) & = & \sum_{k=0}^7 \, \partial_i \, \varphi_k(\blambda) \, f^{(k-5)/3}.
\end{eqnarray}
The Fisher matrix is computed by numerically integrating the expression Eq.~(\ref{eq:metric_v2}) where the derivatives are computed analytically. 
Note that, in this way, the derivatives have to be evaluated only once for one computation of the metric, improving the efficiency of the computation. 

The Fisher matrix computed in physical coordinate system $\blambda \equiv \{M, \eta, \chi\}$ can be transformed to the modified chirp time coordinate system $\btheta \equiv \{\xi_0, \xi_3, \xi_{3S}\}$ in the following way: 
\begin{equation}
{\bm \Gamma}^\prime = \mathbf{J}^\mathrm{T} ~\bm \Gamma ~ \mathbf{J},
\end{equation}
where $\mathbf{J}$ is the Jacobian matrix of coordinate transformation $J_{ik} = \partial \lambda_i/\partial \theta_k$. The Fisher matrix ${\bm \Gamma}^\prime$ in the modified chirp time coordinate system is used to compute the template-space metric using Eq.~(\ref{eq:metricproj}).

\section{Results and discussion}
\label{sec:results}

\begin{figure*}[ht]
\includegraphics[width=0.95\textwidth]{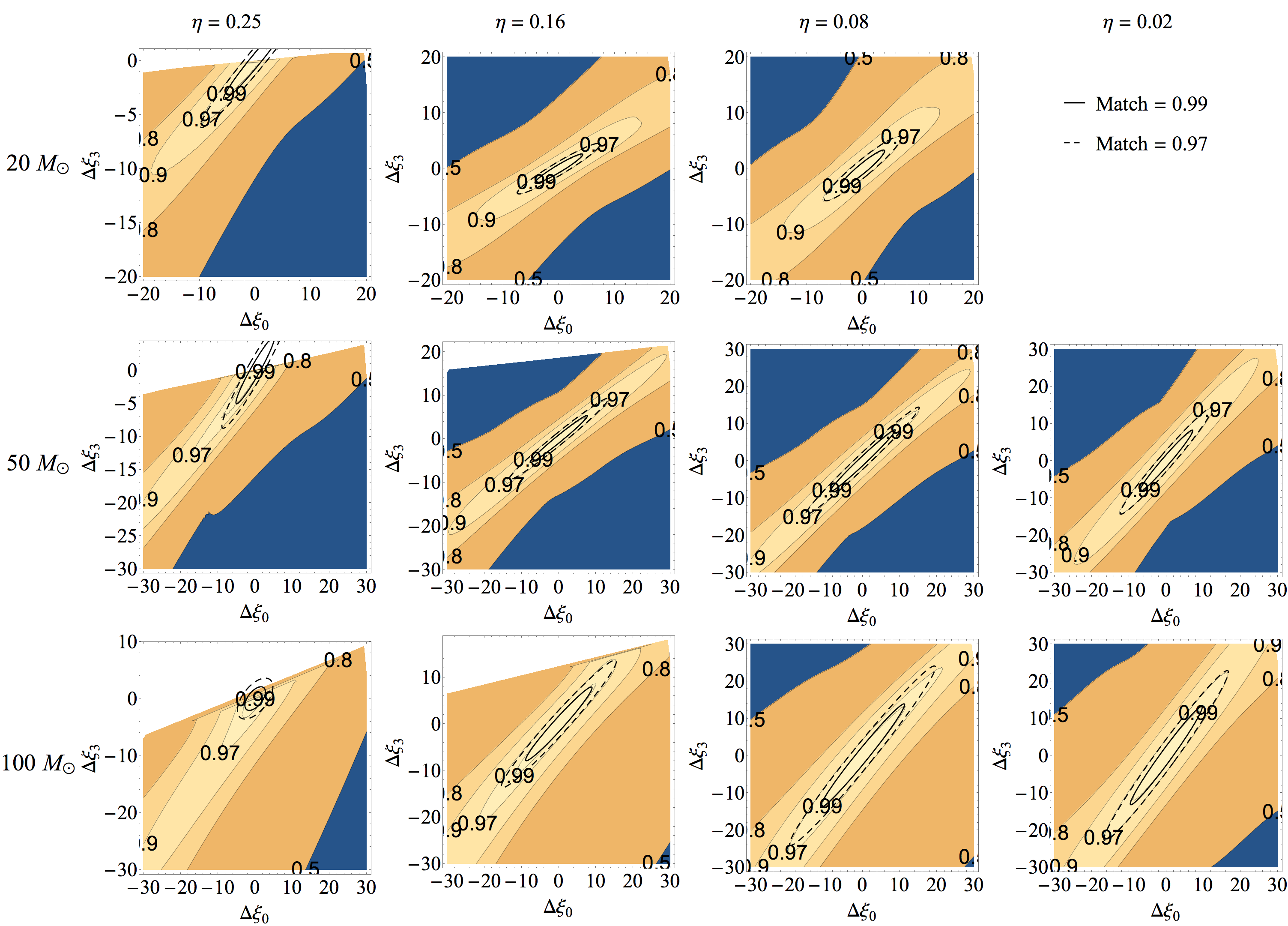}
\caption{Comparison of the match ellipses computed from the non-spinning \imrphenb\ metric (black ellipses) with contours of the ``exact'' match function of \imrphenb\ computed numerically (color contours). The rows correspond to different total masses ($20 M_\odot, 50 M_\odot, 100 M_\odot$) and the columns correspond to different symmetric mass ratios $(0.25, 0.16, 0.08, 0.02)$.  The solid (dashed) black ellipses correspond to a match of $0.99 ~(0.97)$}
\label{fig:MetricApprox2D}
\end{figure*}

\begin{figure*}[ht]
\includegraphics[width=0.95\textwidth]{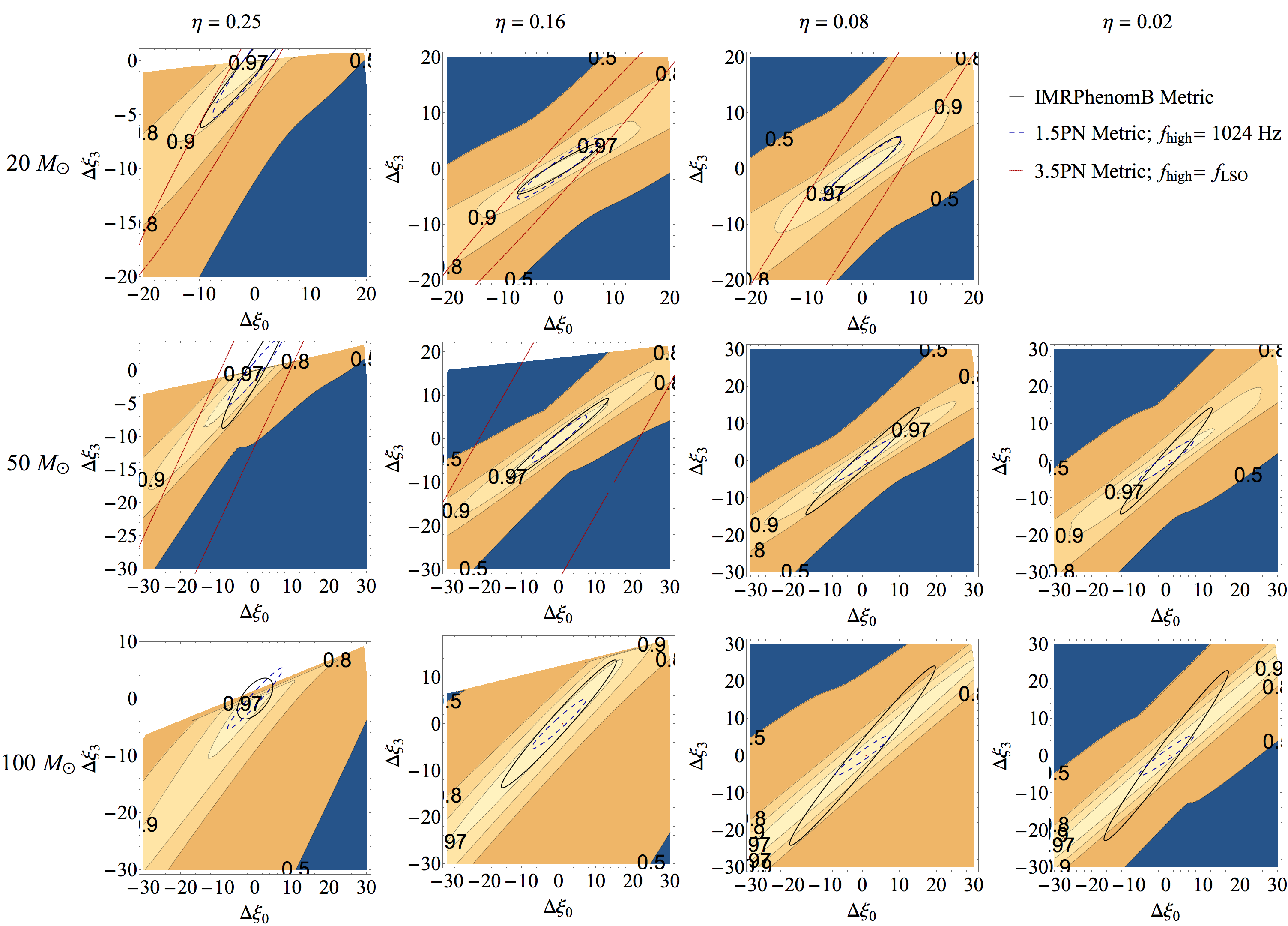}
\caption{Comparison of the match = 0.97 ellipses computed from the non-spinning \imrphenb\ metric (black ellipses) with contours of the numerically-computed ``exact'' match function of the non-spinning \eobnr\ waveforms (color contours). Also shown are the match ellipses from the metric of 3.5PN waveforms truncated at LSO (red ellipses) and the metric of the 1.5PN waveforms truncated at 1024Hz (black dashed ellipses). The rows correspond to different total masses ($20 M_\odot, 50 M_\odot, 100 M_\odot$) and the columns correspond to different symmetric mass ratios $(0.25, 0.16, 0.08, 0.02)$.}
\label{fig:eobnr_comparison}
\end{figure*}

In this section, we discuss the results for the two-dimensional metric of the \emph{nonspinning} \imrphenb\ waveform and the three-dimensional metric of the \emph{non-precessing} \imrphenb\ waveform. Metric calculations have been done assuming \textsc{aLIGOZeroDetHighPower} noise power spectral density of advanced LIGO~\cite{aLIGOPSD} employing a low-frequency cutoff $f_0 = 20$ Hz.

\subsection{Metric of the non-spinning \imrphenb\ waveforms}

The nonspinning \imrphenb\ waveforms,  described by $\{M,\eta,t_0,\phi_0\}$, can be easily obtained by setting the effective spin parameter $\chi=0$ in the waveform discussed in Sec.~\ref{sec:PhenomB}. We then compute the corresponding $4\times4$ Fisher matrix using Eq.~(\ref{eq:Fisher}) and obtain the two dimensional metric by projecting the Fisher matrix orthogonal to $t_0$ and $\phi_0$ using Eq.~(\ref{eq:metricproj}). The quantity \SqrtMet has a large variation over the parameter space while using $(M, \eta)$ as the coordinates (see Fig.~\ref{fig:MetricVariation}). Hence keeping the template placement problem in mind, we use $(\xi_{0}, \xi_{3})$, the PN chirp times coordinates, in the future calculations.

Figure~\ref{fig:MetricApprox2D} compares the ellipses corresponding to constant matches (0.97 and 0.99) obtained semianalytically within the quadratic metric approximation  (black ellipses) with the contours of the match function computed numerically (colored contours). Rows denote different total masses 20\msun, 50\msun\ and 100 \msun\ and the columns denote different symmetric mass ratios 0.25, 0.16, 0.08 and 0.02. It is clear from the figure that the semianalytical and numerical contours agree very well for a match of 0.99 but they differ slightly for a smaller match of 0.97. This is likely due to the inaccuracy of the quadratic approximation to the match function [see Eq.~(\ref{eq:MismatchQuad})] for larger values of parameter differences. The inaccuracy of the metric approximation is the largest for binaries with large, comparable masses, likely due to the fact that they have a small number of GW cycles in the detector band. Despite this, the agreement between the semianalytical and numerical contours is excellent, suggesting that the nonspinning metric computed here can be used for construction of template banks and for multi-detector coincidence tests in searches employing \imrphenb\ templates.

The \imrphenb\ waveform has been calibrated against numerical-relativity simulations with mass ratio $\leq 4$, and is not expected to be faithful towards GW signals from BBHs with large mass ratios~\cite{Ajith:2009bn}. However, several more recent models, in particular the \eobnr\ model, have been constructed and are expected to be more faithful for large mass ratios as well. Having verified that our metric is able to produce an excellent approximation to the match function of the \imrphenb\ waveform, we now investigate the ability of the \imrphenb\ metric, and some variants of the PN metric, to model the match function of the \eobnr\ waveform. 

The results are displayed in Fig.~\ref{fig:eobnr_comparison}. It can be seen that the \imrphenb\ metric produces a reasonable approximation to the size of the 0.97 match contour at all points in the parameter space; but the orientation of the ellipses start to deviate from that of the constant match contours for small mass ratios ($\eta \rightarrow 0$). We also compare the numerical match contours of the \eobnr with two variants of the PN metric: 1) the metric of the restricted PN waveform with 3.5PN accurate phasing terminated at the last stable orbit frequency $f_\mathrm{LSO}$ of the Schwarzschild geometry, and 2) the metric of the restricted 1.5PN waveform terminated at 1024 Hz. The former is the most accurate PN metric available now and the latter is the metric used for two of the previous searches (employing IMR templates) for GWs from high-mass BBHs on the LIGO-Virgo data~\cite{Aasi:2012rja,Abadie:2011kd}. It is evident that the 3.5PN metric overestimates the size of the match contour significantly. The overestimation becomes severe as we go to higher mass systems. Using this metric in the template placement will cause severe under-coverage of the parameter space. Surprisingly, the metric of 1.5PN waveform terminated at 1024 Hz provides a reasonable approximation to the \eobnr\ match contours at low masses. However, in the high-mass regime $(M \gtrsim 50 M_\odot)$, the analytical match ellipses computed using this metric are significantly smaller than the exact match contours. This will cause significant over-coverage of the template bank when used in template placement. However, we note that it might be possible to improve the agreement of this metric (which, we believe is accidental) by tuning the termination frequency. 

\begin{figure*}[t]
\includegraphics[width=1\textwidth]{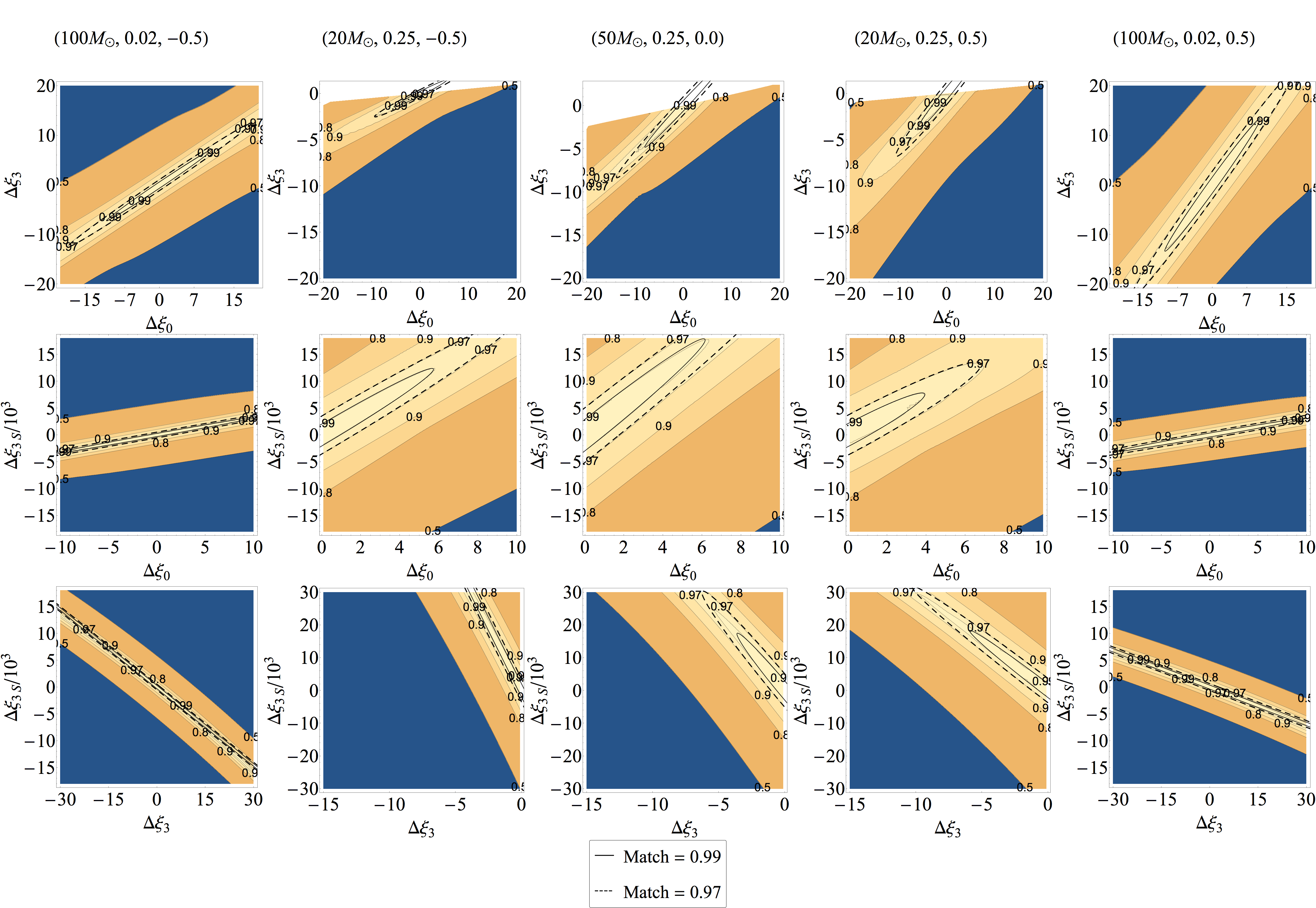}
\caption{Comparison of the match ellipses computed from the non-precessing spin \imrphenb\ metric (black ellipses) with contours of the ``exact'' match function of \imrphenb\ computed numerically (color contours). The total mass, symmetric mass ratio, and the effective spin parameter ($M, \eta, \chi$) corresponding to point in the parameter space relative to which the match function is computed is shown on the top of each column. The different rows correspond to two-dimensional slices of these contours in the $\Delta \xi_0 - \Delta \xi_3$ plane (top row),  $\Delta \xi_0 - \Delta \xi_{3S}$ plane (middle row) and $\Delta \xi_3 - \Delta \xi_{3S}$ plane (bottom row). The solid black ellipses correspond to a match of $0.99$ and dashed black ellipses correspond to a match of $0.97$.}
\label{fig:MetricApprox3D}
\end{figure*}

\subsection{Metric of the non-precessing spin \imrphenb\ waveforms}

The three-dimensional metric corresponding to the two mass parameters and the effective spin parameter can be computed starting from a $5\times5$ Fisher matrix corresponding to $\{\xi_0,\xi_3,\xi_{3S}, t_0,\phi_0\}$, where $\xi_k$ represent the new coordinates introduced in Sec.~\ref{sec:NewCord}. Projection of the Fisher matrix orthogonal to ($t_0,\phi_0$) will give the three dimensional \imrphenb\ metric for non-precessing binaries. Ellipses corresponding to the two-dimensional slices of the three-dimensional ellipsoid is shown in Fig.~\ref{fig:MetricApprox3D}, which are compared against the two-dimensional slices of the contours of the three-dimensional match function computed numerically for \imrphenb\ waveform. Each column specifies the total mass, symmetric mass ratio and effective spin parameter that are considered. The solid black ellipses represent the contours of match 0.99 and the dashed black ones correspond to a match of 0.97. We find excellent agreement between the ellipses obtained from metric and the numerical contours for a match of 0.99. The agreement is not so good for a match of 0.97 as it is for a match of 0.99. Such a disagreement we have already encountered in the case of two dimensional ellipses and is argued to be due to the break down of the (quadratic) metric approximation to the match function. However, in general, the agreement between the analytically computed match function from the metric and the numerical exact match contours is excellent.

\section{Conclusion and future directions}
\label{sec:conclusions}
In this paper we computed the template-space metric of the inspiral, merger, ringdown gravitational waveform family \imrphenb. We proposed a coordinate system (which is an adaptation of the PN chirp time coordinates) in which the metric is slowly varying over the parameter space of interest. The semianalytical match function computed using the metric has excellent agreement with the exact match function of the \imrphenb\ waveform computed numerically. In addition, we have also shown that the analytical match function computed using the \imrphenb\ metric agrees well with the exact match function of other IMR waveform families, such as the \eobnr\ over the entire parameter space of interest. This will potentially allow us to use the \imrphenb\ metric in the searches for GWs from BBHs that employ more accurate IMR waveform families as well (the ones that are already available and those under development). There is ongoing work to employ this metric in the construction of IMR template banks (primarily using stochastic placement methods) as well as in multi-detector coincidence tests~\cite{Gaur:2015aa}. We anticipate that this will contribute to improving the sensitivity of searches for GWs from BBHs in the advanced detector era. 

We note that there is an alternative proposal to use ``exact coincidence'' criterion between multi-detector triggers, whereby the same template bank is used in multiple detectors and two triggers are considered coincident only if they are detected with exactly the same parameters (see, e.g.,~\cite{Privitera:2013xza}). While this method has its own advantages, it is quite possible that a given signal is ascribed different parameters in two detectors due to the presence of noise, potentially missing a coincidence. Thus, designing a coincidence criterion based on the real metric of the templates is likely to be advantageous. However, only an apples-to-apples comparison can quantify the advantage or disadvantage of these methods, and we leave this as future work. 


\acknowledgments
We thank Ian Harry and Michael P\"urrer for useful comments on the manuscript and Sukanta Bose, Thomas Dent, Bala Iyer, Stephen Privitera and B S Sathyaprakash for useful discussions. This work was initiated during the ICTS Program on Numerical relativity organized by the International Center for Theoretical Sciences (ICTS), Bangalore during June-July 2013. CK thanks ICTS for support during the S.N. Bhatt Memorial Excellence Fellowship Program 2014. CK and KGA thank ICTS for hospitality. CK and KGA were partially funded by a grant from Infosys Foundation. PA's research was supported by a Ramanujan Fellowship from the Science and Engineering Research Board (SERB), India, the SERB FastTrack fellowship SR/FTP/PS-191/2012, and by the AIRBUS Group Corporate Foundation through a chair in ``Mathematics of Complex Systems'' at ICTS, and by the Max Planck Society and the Department of Science and Technology, India through a Max Planck Partner Group at ICTS. Computations were performed using the ICTS computing cluster Mowgli. This paper has the LIGO document number LIGO-P1400252-v3.

\bibliography{Metric}

\begin{thebibliography}{63}
\expandafter\ifx\csname natexlab\endcsname\relax\def\natexlab#1{#1}\fi
\expandafter\ifx\csname bibnamefont\endcsname\relax
  \def\bibnamefont#1{#1}\fi
\expandafter\ifx\csname bibfnamefont\endcsname\relax
  \def\bibfnamefont#1{#1}\fi
\expandafter\ifx\csname citenamefont\endcsname\relax
  \def\citenamefont#1{#1}\fi
\expandafter\ifx\csname url\endcsname\relax
  \def\url#1{\texttt{#1}}\fi
\expandafter\ifx\csname urlprefix\endcsname\relax\def\urlprefix{URL }\fi
\providecommand{\bibinfo}[2]{#2}
\providecommand{\eprint}[2][]{\url{#2}}

\bibitem[{\citenamefont{Harry and the LIGO
  Scientific~Collaboration}(2010)}]{0264-9381-27-8-084006}
\bibinfo{author}{\bibfnamefont{G.~M.} \bibnamefont{Harry}} \bibnamefont{and}
  \bibinfo{author}{\bibnamefont{the LIGO Scientific~Collaboration}},
  \bibinfo{journal}{Classical and Quantum Gravity}
  \textbf{\bibinfo{volume}{27}}, \bibinfo{pages}{084006}
  (\bibinfo{year}{2010}),
  \urlprefix\url{http://stacks.iop.org/0264-9381/27/i=8/a=084006}.

\bibitem[{\citenamefont{{{The} {Virgo} {Collaboration}}}(2009)}]{AdvVirgo:2009}
\bibinfo{author}{\bibnamefont{{{The} {Virgo} {Collaboration}}}}
  (\bibinfo{year}{2009}), \bibinfo{note}{{Virgo} Technical Document
  VIR-027A-09},
  \urlprefix\url{https://tds.ego-gw.it/itf/tds/file.php?callFile=VIR-0027A-09.pdf}.

\bibitem[{\citenamefont{Aasi et~al.}(2013{\natexlab{a}})}]{Aasi:2013wya}
\bibinfo{author}{\bibfnamefont{J.}~\bibnamefont{Aasi}} \bibnamefont{et~al.}
  (\bibinfo{collaboration}{LIGO Scientific Collaboration, Virgo Collaboration})
  (\bibinfo{year}{2013}{\natexlab{a}}), \eprint{1304.0670}.

\bibitem[{\citenamefont{Somiya}(2012)}]{Somiya-Kagra:2012}
\bibinfo{author}{\bibfnamefont{K.}~\bibnamefont{Somiya}},
  \bibinfo{journal}{Classical and Quantum Gravity}
  \textbf{\bibinfo{volume}{29}}, \bibinfo{pages}{124007}
  (\bibinfo{year}{2012}),
  \urlprefix\url{http://stacks.iop.org/0264-9381/29/i=12/a=124007}.

\bibitem[{\citenamefont{Iyer et~al.}(2011)\citenamefont{Iyer, Souradeep,
  Unnikrishnan, Dhurandhar, Raja, and Sengupta}}]{LIGOIndiaProposal:2011}
\bibinfo{author}{\bibfnamefont{B.}~\bibnamefont{Iyer}},
  \bibinfo{author}{\bibfnamefont{T.}~\bibnamefont{Souradeep}},
  \bibinfo{author}{\bibfnamefont{C.}~\bibnamefont{Unnikrishnan}},
  \bibinfo{author}{\bibfnamefont{S.}~\bibnamefont{Dhurandhar}},
  \bibinfo{author}{\bibfnamefont{S.}~\bibnamefont{Raja}}, \bibnamefont{and}
  \bibinfo{author}{\bibfnamefont{A.}~\bibnamefont{Sengupta}}
  (\bibinfo{year}{2011}), \bibinfo{note}{{LIGO} Technical Document
  LIGO-M1100296-v2}, \urlprefix\url{https://dcc.ligo.org/LIGO-M1100296/public}.

\bibitem[{\citenamefont{{Abadie} et~al.}(2010)\citenamefont{{Abadie}, {Abbott},
  {Abbott}, {Abernathy}, {Accadia}, {Acernese}, {Adams}, {Adhikari}, {Ajith},
  {Allen} et~al.}}]{2010CQGra..27q3001A}
\bibinfo{author}{\bibfnamefont{J.}~\bibnamefont{{Abadie}}},
  \bibinfo{author}{\bibfnamefont{B.~P.} \bibnamefont{{Abbott}}},
  \bibinfo{author}{\bibfnamefont{R.}~\bibnamefont{{Abbott}}},
  \bibinfo{author}{\bibfnamefont{M.}~\bibnamefont{{Abernathy}}},
  \bibinfo{author}{\bibfnamefont{T.}~\bibnamefont{{Accadia}}},
  \bibinfo{author}{\bibfnamefont{F.}~\bibnamefont{{Acernese}}},
  \bibinfo{author}{\bibfnamefont{C.}~\bibnamefont{{Adams}}},
  \bibinfo{author}{\bibfnamefont{R.}~\bibnamefont{{Adhikari}}},
  \bibinfo{author}{\bibfnamefont{P.}~\bibnamefont{{Ajith}}},
  \bibinfo{author}{\bibfnamefont{B.}~\bibnamefont{{Allen}}},
  \bibnamefont{et~al.}, \bibinfo{journal}{Classical and Quantum Gravity}
  \textbf{\bibinfo{volume}{27}}, \bibinfo{eid}{173001} (\bibinfo{year}{2010}),
  \eprint{1003.2480}.

\bibitem[{\citenamefont{Sathyaprakash and Schutz}(2009)}]{SathyaSchutzLivRev09}
\bibinfo{author}{\bibfnamefont{B.}~\bibnamefont{Sathyaprakash}}
  \bibnamefont{and} \bibinfo{author}{\bibfnamefont{B.}~\bibnamefont{Schutz}},
  \bibinfo{journal}{Living Rev.Rel.} \textbf{\bibinfo{volume}{12}},
  \bibinfo{pages}{2} (\bibinfo{year}{2009}), \eprint{arXiv:0903.0338}.

\bibitem[{\citenamefont{Blanchet}(2006{\natexlab{a}})}]{Blanchet:LivRev}
\bibinfo{author}{\bibfnamefont{L.}~\bibnamefont{Blanchet}},
  \bibinfo{journal}{Living Rev. Relativity} \textbf{\bibinfo{volume}{9}}
  (\bibinfo{year}{2006}{\natexlab{a}}), \eprint{gr-qc/0202016},
  \urlprefix\url{http://www.livingreviews.org/lrr-2006-4}.

\bibitem[{\citenamefont{Centrella et~al.}(2010)\citenamefont{Centrella, Baker,
  Kelly, and van Meter}}]{Centrella:2010mx}
\bibinfo{author}{\bibfnamefont{J.~M.} \bibnamefont{Centrella}},
  \bibinfo{author}{\bibfnamefont{J.~G.} \bibnamefont{Baker}},
  \bibinfo{author}{\bibfnamefont{B.~J.} \bibnamefont{Kelly}}, \bibnamefont{and}
  \bibinfo{author}{\bibfnamefont{J.~R.} \bibnamefont{van Meter}},
  \bibinfo{journal}{Rev.Mod.Phys.} \textbf{\bibinfo{volume}{82}},
  \bibinfo{pages}{3069} (\bibinfo{year}{2010}), \eprint{1010.5260}.

\bibitem[{\citenamefont{Berti et~al.}(2009)\citenamefont{Berti, Cardoso, and
  Starinets}}]{Berti:2009kk}
\bibinfo{author}{\bibfnamefont{E.}~\bibnamefont{Berti}},
  \bibinfo{author}{\bibfnamefont{V.}~\bibnamefont{Cardoso}}, \bibnamefont{and}
  \bibinfo{author}{\bibfnamefont{A.~O.} \bibnamefont{Starinets}},
  \bibinfo{journal}{Class.Quant.Grav.} \textbf{\bibinfo{volume}{26}},
  \bibinfo{pages}{163001} (\bibinfo{year}{2009}), \eprint{0905.2975}.

\bibitem[{\citenamefont{Babak et~al.}(2013)\citenamefont{Babak, Biswas, Brady,
  Brown, Cannon et~al.}}]{Babak:2012zx}
\bibinfo{author}{\bibfnamefont{S.}~\bibnamefont{Babak}},
  \bibinfo{author}{\bibfnamefont{R.}~\bibnamefont{Biswas}},
  \bibinfo{author}{\bibfnamefont{P.}~\bibnamefont{Brady}},
  \bibinfo{author}{\bibfnamefont{D.}~\bibnamefont{Brown}},
  \bibinfo{author}{\bibfnamefont{K.}~\bibnamefont{Cannon}},
  \bibnamefont{et~al.}, \bibinfo{journal}{Phys.Rev.}
  \textbf{\bibinfo{volume}{D87}}, \bibinfo{pages}{024033}
  (\bibinfo{year}{2013}), \eprint{1208.3491}.

\bibitem[{\citenamefont{Aasi et~al.}(2013{\natexlab{b}})}]{Aasi:2012rja}
\bibinfo{author}{\bibfnamefont{J.}~\bibnamefont{Aasi}} \bibnamefont{et~al.}
  (\bibinfo{collaboration}{LIGO Scientific Collaboration, Virgo
  Collaboration}), \bibinfo{journal}{Phys.Rev.} \textbf{\bibinfo{volume}{D87}},
  \bibinfo{pages}{022002} (\bibinfo{year}{2013}{\natexlab{b}}),
  \eprint{1209.6533}.

\bibitem[{\citenamefont{Abadie et~al.}(2011)}]{Abadie:2011kd}
\bibinfo{author}{\bibfnamefont{J.}~\bibnamefont{Abadie}} \bibnamefont{et~al.}
  (\bibinfo{collaboration}{LIGO Scientific Collaboration and the Virgo
  Collaboration, Virgo Collaboration}), \bibinfo{journal}{Phys.Rev.}
  \textbf{\bibinfo{volume}{D83}}, \bibinfo{pages}{122005}
  (\bibinfo{year}{2011}), \eprint{1102.3781}.

\bibitem[{\citenamefont{Blanchet}(2006{\natexlab{b}})}]{Bliving}
\bibinfo{author}{\bibfnamefont{L.}~\bibnamefont{Blanchet}},
  \bibinfo{journal}{Living Rev. Rel.} \textbf{\bibinfo{volume}{9}},
  \bibinfo{pages}{4} (\bibinfo{year}{2006}{\natexlab{b}}),
  \eprint{arXiv:1310.1528}.

\bibitem[{\citenamefont{Sasaki and Tagoshi}(2003)}]{TSLivRev03}
\bibinfo{author}{\bibfnamefont{M.}~\bibnamefont{Sasaki}} \bibnamefont{and}
  \bibinfo{author}{\bibfnamefont{H.}~\bibnamefont{Tagoshi}},
  \bibinfo{journal}{Living Rev. Rel.} \textbf{\bibinfo{volume}{6}},
  \bibinfo{pages}{6} (\bibinfo{year}{2003}), \eprint{gr-qc/0306120}.

\bibitem[{\citenamefont{Abbott et~al.}(2008{\natexlab{a}})}]{Abbott:2007xi}
\bibinfo{author}{\bibfnamefont{B.}~\bibnamefont{Abbott}} \bibnamefont{et~al.}
  (\bibinfo{collaboration}{LIGO Scientific Collaboration}),
  \bibinfo{journal}{Phys.Rev.} \textbf{\bibinfo{volume}{D77}},
  \bibinfo{pages}{062002} (\bibinfo{year}{2008}{\natexlab{a}}),
  \eprint{0704.3368}.

\bibitem[{\citenamefont{Abbott et~al.}(2008{\natexlab{b}})}]{Abbott:2007ai}
\bibinfo{author}{\bibfnamefont{B.}~\bibnamefont{Abbott}} \bibnamefont{et~al.}
  (\bibinfo{collaboration}{LIGO Scientific Collaboration}),
  \bibinfo{journal}{Phys.Rev.} \textbf{\bibinfo{volume}{D78}},
  \bibinfo{pages}{042002} (\bibinfo{year}{2008}{\natexlab{b}}),
  \eprint{0712.2050}.

\bibitem[{\citenamefont{Abbott et~al.}(2009)}]{Abbott:2009km}
\bibinfo{author}{\bibfnamefont{B.}~\bibnamefont{Abbott}} \bibnamefont{et~al.}
  (\bibinfo{collaboration}{LIGO Scientific Collaboration}),
  \bibinfo{journal}{Phys.Rev.} \textbf{\bibinfo{volume}{D80}},
  \bibinfo{pages}{062001} (\bibinfo{year}{2009}), \eprint{0905.1654}.

\bibitem[{\citenamefont{Aasi et~al.}(2014{\natexlab{a}})}]{Aasi:2014bqj}
\bibinfo{author}{\bibfnamefont{J.}~\bibnamefont{Aasi}} \bibnamefont{et~al.}
  (\bibinfo{collaboration}{LIGO Scientific Collaboration, VIRGO
  Collaboration}), \bibinfo{journal}{Phys.Rev.} \textbf{\bibinfo{volume}{D89}},
  \bibinfo{pages}{102006} (\bibinfo{year}{2014}{\natexlab{a}}),
  \eprint{1403.5306}.

\bibitem[{\citenamefont{Acernese et~al.}(2009)}]{VirgoC7}
\bibinfo{author}{\bibfnamefont{F.}~\bibnamefont{Acernese}} \bibnamefont{et~al.}
  (\bibinfo{collaboration}{Virgo Collaboration}), \bibinfo{journal}{Class.
  Quant. Grav.} \textbf{\bibinfo{volume}{26}}, \bibinfo{pages}{085009}
  (\bibinfo{year}{2009}), \eprint{0812.4870}.

\bibitem[{\citenamefont{Abadie et~al.}(2012)}]{Virgo:2012aa}
\bibinfo{author}{\bibfnamefont{J.}~\bibnamefont{Abadie}} \bibnamefont{et~al.}
  (\bibinfo{collaboration}{LIGO Scientific Collaboration, Virgo
  Collaboration}), \bibinfo{journal}{Phys.Rev.} \textbf{\bibinfo{volume}{D85}},
  \bibinfo{pages}{102004} (\bibinfo{year}{2012}), \eprint{1201.5999}.

\bibitem[{\citenamefont{Aasi et~al.}(2014{\natexlab{b}})}]{Aasi:2014iwa}
\bibinfo{author}{\bibfnamefont{J.}~\bibnamefont{Aasi}} \bibnamefont{et~al.}
  (\bibinfo{collaboration}{The LIGO Scientific Collaboration, the Virgo
  Collaboration}), \bibinfo{journal}{Phys.Rev.} \textbf{\bibinfo{volume}{D89}},
  \bibinfo{pages}{122003} (\bibinfo{year}{2014}{\natexlab{b}}),
  \eprint{1404.2199}.

\bibitem[{\citenamefont{Pretorius}(2005)}]{Pretorius:2005gq}
\bibinfo{author}{\bibfnamefont{F.}~\bibnamefont{Pretorius}},
  \bibinfo{journal}{Phys. Rev. Lett.} \textbf{\bibinfo{volume}{95}},
  \bibinfo{pages}{121101} (\bibinfo{year}{2005}), \eprint{gr-qc/0507014}.

\bibitem[{\citenamefont{Campanelli et~al.}(2006)\citenamefont{Campanelli,
  Lousto, Marronetti, and Zlochower}}]{Campanelli:2005dd}
\bibinfo{author}{\bibfnamefont{M.}~\bibnamefont{Campanelli}},
  \bibinfo{author}{\bibfnamefont{C.~O.} \bibnamefont{Lousto}},
  \bibinfo{author}{\bibfnamefont{P.}~\bibnamefont{Marronetti}},
  \bibnamefont{and}
  \bibinfo{author}{\bibfnamefont{Y.}~\bibnamefont{Zlochower}},
  \bibinfo{journal}{Phys. Rev. Lett.} \textbf{\bibinfo{volume}{96}},
  \bibinfo{pages}{111101} (\bibinfo{year}{2006}), \eprint{gr-qc/0511048}.

\bibitem[{\citenamefont{Baker et~al.}(2006)}]{Baker05a}
\bibinfo{author}{\bibfnamefont{J.~G.} \bibnamefont{Baker}}
  \bibnamefont{et~al.}, \bibinfo{journal}{Phys. Rev. Lett.}
  \textbf{\bibinfo{volume}{96}}, \bibinfo{pages}{111102}
  (\bibinfo{year}{2006}).

\bibitem[{\citenamefont{Ajith et~al.}(2012)\citenamefont{Ajith, Boyle, Brown,
  Br{\"u}gmann, Buchman et~al.}}]{Ajith:2012az}
\bibinfo{author}{\bibfnamefont{P.}~\bibnamefont{Ajith}},
  \bibinfo{author}{\bibfnamefont{M.}~\bibnamefont{Boyle}},
  \bibinfo{author}{\bibfnamefont{D.~A.} \bibnamefont{Brown}},
  \bibinfo{author}{\bibfnamefont{B.}~\bibnamefont{Br{\"u}gmann}},
  \bibinfo{author}{\bibfnamefont{L.~T.} \bibnamefont{Buchman}},
  \bibnamefont{et~al.}, \bibinfo{journal}{Class.Quant.Grav.}
  \textbf{\bibinfo{volume}{29}}, \bibinfo{pages}{124001}
  (\bibinfo{year}{2012}), \eprint{1201.5319}.

\bibitem[{SXS()}]{SXS-Catalog}
\bibinfo{note}{SXS Gravitational Waveform Database},
  \urlprefix\url{http://www.black-holes.org/waveforms/}.

\bibitem[{\citenamefont{Mroue et~al.}(2013)\citenamefont{Mroue, Scheel,
  Szilagyi, Pfeiffer, Boyle et~al.}}]{Mroue:2013xna}
\bibinfo{author}{\bibfnamefont{A.~H.} \bibnamefont{Mroue}},
  \bibinfo{author}{\bibfnamefont{M.~A.} \bibnamefont{Scheel}},
  \bibinfo{author}{\bibfnamefont{B.}~\bibnamefont{Szilagyi}},
  \bibinfo{author}{\bibfnamefont{H.~P.} \bibnamefont{Pfeiffer}},
  \bibinfo{author}{\bibfnamefont{M.}~\bibnamefont{Boyle}},
  \bibnamefont{et~al.}, \bibinfo{journal}{Phys.Rev.Lett.}
  \textbf{\bibinfo{volume}{111}}, \bibinfo{pages}{241104}
  (\bibinfo{year}{2013}), \eprint{1304.6077}.

\bibitem[{\citenamefont{Buonanno et~al.}(2007)}]{Buonanno:2007pf}
\bibinfo{author}{\bibfnamefont{A.}~\bibnamefont{Buonanno}}
  \bibnamefont{et~al.}, \bibinfo{journal}{Phys. Rev.}
  \textbf{\bibinfo{volume}{D76}}, \bibinfo{pages}{104049}
  (\bibinfo{year}{2007}), \eprint{0706.3732}.

\bibitem[{\citenamefont{Pan et~al.}(2011)\citenamefont{Pan, Buonanno, Boyle,
  Buchman, Kidder et~al.}}]{Pan:2011gk}
\bibinfo{author}{\bibfnamefont{Y.}~\bibnamefont{Pan}},
  \bibinfo{author}{\bibfnamefont{A.}~\bibnamefont{Buonanno}},
  \bibinfo{author}{\bibfnamefont{M.}~\bibnamefont{Boyle}},
  \bibinfo{author}{\bibfnamefont{L.~T.} \bibnamefont{Buchman}},
  \bibinfo{author}{\bibfnamefont{L.~E.} \bibnamefont{Kidder}},
  \bibnamefont{et~al.}, \bibinfo{journal}{Phys.Rev.}
  \textbf{\bibinfo{volume}{D84}}, \bibinfo{pages}{124052}
  (\bibinfo{year}{2011}), \eprint{1106.1021}.

\bibitem[{\citenamefont{Taracchini et~al.}(2014)\citenamefont{Taracchini,
  Buonanno, Pan, Hinderer, Boyle et~al.}}]{Taracchini:2013rva}
\bibinfo{author}{\bibfnamefont{A.}~\bibnamefont{Taracchini}},
  \bibinfo{author}{\bibfnamefont{A.}~\bibnamefont{Buonanno}},
  \bibinfo{author}{\bibfnamefont{Y.}~\bibnamefont{Pan}},
  \bibinfo{author}{\bibfnamefont{T.}~\bibnamefont{Hinderer}},
  \bibinfo{author}{\bibfnamefont{M.}~\bibnamefont{Boyle}},
  \bibnamefont{et~al.}, \bibinfo{journal}{Phys.Rev.}
  \textbf{\bibinfo{volume}{D89}}, \bibinfo{pages}{061502}
  (\bibinfo{year}{2014}), \eprint{1311.2544}.

\bibitem[{\citenamefont{Pan et~al.}(2014)\citenamefont{Pan, Buonanno,
  Taracchini, Kidder, Mrou{\'e} et~al.}}]{Pan:2013rra}
\bibinfo{author}{\bibfnamefont{Y.}~\bibnamefont{Pan}},
  \bibinfo{author}{\bibfnamefont{A.}~\bibnamefont{Buonanno}},
  \bibinfo{author}{\bibfnamefont{A.}~\bibnamefont{Taracchini}},
  \bibinfo{author}{\bibfnamefont{L.~E.} \bibnamefont{Kidder}},
  \bibinfo{author}{\bibfnamefont{A.~H.} \bibnamefont{Mrou{\'e}}},
  \bibnamefont{et~al.}, \bibinfo{journal}{Phys.Rev.}
  \textbf{\bibinfo{volume}{D89}}, \bibinfo{pages}{084006}
  (\bibinfo{year}{2014}), \eprint{1307.6232}.

\bibitem[{\citenamefont{Taracchini et~al.}(2012)\citenamefont{Taracchini, Pan,
  Buonanno, Barausse, Boyle et~al.}}]{Taracchini:2012ig}
\bibinfo{author}{\bibfnamefont{A.}~\bibnamefont{Taracchini}},
  \bibinfo{author}{\bibfnamefont{Y.}~\bibnamefont{Pan}},
  \bibinfo{author}{\bibfnamefont{A.}~\bibnamefont{Buonanno}},
  \bibinfo{author}{\bibfnamefont{E.}~\bibnamefont{Barausse}},
  \bibinfo{author}{\bibfnamefont{M.}~\bibnamefont{Boyle}},
  \bibnamefont{et~al.}, \bibinfo{journal}{Phys.Rev.}
  \textbf{\bibinfo{volume}{D86}}, \bibinfo{pages}{024011}
  (\bibinfo{year}{2012}), \eprint{1202.0790}.

\bibitem[{\citenamefont{Damour et~al.}(2008{\natexlab{a}})\citenamefont{Damour,
  Nagar, Dorband, Pollney, and Rezzolla}}]{Damour:2007vq}
\bibinfo{author}{\bibfnamefont{T.}~\bibnamefont{Damour}},
  \bibinfo{author}{\bibfnamefont{A.}~\bibnamefont{Nagar}},
  \bibinfo{author}{\bibfnamefont{E.~N.} \bibnamefont{Dorband}},
  \bibinfo{author}{\bibfnamefont{D.}~\bibnamefont{Pollney}}, \bibnamefont{and}
  \bibinfo{author}{\bibfnamefont{L.}~\bibnamefont{Rezzolla}},
  \bibinfo{journal}{Phys.Rev.} \textbf{\bibinfo{volume}{D77}},
  \bibinfo{pages}{084017} (\bibinfo{year}{2008}{\natexlab{a}}),
  \eprint{0712.3003}.

\bibitem[{\citenamefont{Damour et~al.}(2008{\natexlab{b}})\citenamefont{Damour,
  Nagar, Hannam, Husa, and Bruegmann}}]{Damour:2008te}
\bibinfo{author}{\bibfnamefont{T.}~\bibnamefont{Damour}},
  \bibinfo{author}{\bibfnamefont{A.}~\bibnamefont{Nagar}},
  \bibinfo{author}{\bibfnamefont{M.}~\bibnamefont{Hannam}},
  \bibinfo{author}{\bibfnamefont{S.}~\bibnamefont{Husa}}, \bibnamefont{and}
  \bibinfo{author}{\bibfnamefont{B.}~\bibnamefont{Bruegmann}},
  \bibinfo{journal}{Phys.Rev.} \textbf{\bibinfo{volume}{D78}},
  \bibinfo{pages}{044039} (\bibinfo{year}{2008}{\natexlab{b}}),
  \eprint{0803.3162}.

\bibitem[{\citenamefont{Damour and Nagar}(2009)}]{Damour:2009kr}
\bibinfo{author}{\bibfnamefont{T.}~\bibnamefont{Damour}} \bibnamefont{and}
  \bibinfo{author}{\bibfnamefont{A.}~\bibnamefont{Nagar}},
  \bibinfo{journal}{Phys.Rev.} \textbf{\bibinfo{volume}{D79}},
  \bibinfo{pages}{081503} (\bibinfo{year}{2009}), \eprint{0902.0136}.

\bibitem[{\citenamefont{Damour et~al.}(2013)\citenamefont{Damour, Nagar, and
  Bernuzzi}}]{Damour:2012ky}
\bibinfo{author}{\bibfnamefont{T.}~\bibnamefont{Damour}},
  \bibinfo{author}{\bibfnamefont{A.}~\bibnamefont{Nagar}}, \bibnamefont{and}
  \bibinfo{author}{\bibfnamefont{S.}~\bibnamefont{Bernuzzi}},
  \bibinfo{journal}{Phys.Rev.} \textbf{\bibinfo{volume}{D87}},
  \bibinfo{pages}{084035} (\bibinfo{year}{2013}), \eprint{1212.4357}.

\bibitem[{\citenamefont{Damour and Nagar}(2014)}]{Damour:2014sva}
\bibinfo{author}{\bibfnamefont{T.}~\bibnamefont{Damour}} \bibnamefont{and}
  \bibinfo{author}{\bibfnamefont{A.}~\bibnamefont{Nagar}},
  \bibinfo{journal}{Phys.Rev.} \textbf{\bibinfo{volume}{D90}},
  \bibinfo{pages}{044018} (\bibinfo{year}{2014}), \eprint{1406.6913}.

\bibitem[{\citenamefont{Ajith et~al.}(2007)}]{Ajith:2007qp}
\bibinfo{author}{\bibfnamefont{P.}~\bibnamefont{Ajith}} \bibnamefont{et~al.},
  \bibinfo{journal}{Class. Quant. Grav.} \textbf{\bibinfo{volume}{24}},
  \bibinfo{pages}{S689} (\bibinfo{year}{2007}).

\bibitem[{\citenamefont{Ajith et~al.}(2011)\citenamefont{Ajith, Hannam, Husa,
  Chen, Br\"ugmann, Dorband, M\"uller, Ohme, Pollney, Reisswig
  et~al.}}]{Ajith:2009bn}
\bibinfo{author}{\bibfnamefont{P.}~\bibnamefont{Ajith}},
  \bibinfo{author}{\bibfnamefont{M.}~\bibnamefont{Hannam}},
  \bibinfo{author}{\bibfnamefont{S.}~\bibnamefont{Husa}},
  \bibinfo{author}{\bibfnamefont{Y.}~\bibnamefont{Chen}},
  \bibinfo{author}{\bibfnamefont{B.}~\bibnamefont{Br\"ugmann}},
  \bibinfo{author}{\bibfnamefont{N.}~\bibnamefont{Dorband}},
  \bibinfo{author}{\bibfnamefont{D.}~\bibnamefont{M\"uller}},
  \bibinfo{author}{\bibfnamefont{F.}~\bibnamefont{Ohme}},
  \bibinfo{author}{\bibfnamefont{D.}~\bibnamefont{Pollney}},
  \bibinfo{author}{\bibfnamefont{C.}~\bibnamefont{Reisswig}},
  \bibnamefont{et~al.}, \bibinfo{journal}{Phys. Rev. Lett.}
  \textbf{\bibinfo{volume}{106}}, \bibinfo{pages}{241101}
  (\bibinfo{year}{2011}), \eprint{0909.2867}.

\bibitem[{\citenamefont{Ajith}(2008)}]{Ajith:2007xh}
\bibinfo{author}{\bibfnamefont{P.}~\bibnamefont{Ajith}},
  \bibinfo{journal}{Class. Quant. Grav.} \textbf{\bibinfo{volume}{25}},
  \bibinfo{pages}{114033} (\bibinfo{year}{2008}).

\bibitem[{\citenamefont{Ajith et~al.}(2008)}]{Ajith:2007kx}
\bibinfo{author}{\bibfnamefont{P.}~\bibnamefont{Ajith}} \bibnamefont{et~al.},
  \bibinfo{journal}{Phys. Rev.} \textbf{\bibinfo{volume}{D77}},
  \bibinfo{pages}{104017} (\bibinfo{year}{2008}), \eprint{0710.2335}.

\bibitem[{\citenamefont{Santamaria et~al.}(2010)}]{Santamaria:2010yb}
\bibinfo{author}{\bibfnamefont{L.}~\bibnamefont{Santamaria}}
  \bibnamefont{et~al.}, \bibinfo{journal}{Phys. Rev.}
  \textbf{\bibinfo{volume}{D82}}, \bibinfo{pages}{064016}
  (\bibinfo{year}{2010}), \eprint{1005.3306}.

\bibitem[{\citenamefont{Hannam et~al.}(2014)\citenamefont{Hannam, Schmidt,
  Boh{\'e}, Haegel, Husa et~al.}}]{Hannam:2013oca}
\bibinfo{author}{\bibfnamefont{M.}~\bibnamefont{Hannam}},
  \bibinfo{author}{\bibfnamefont{P.}~\bibnamefont{Schmidt}},
  \bibinfo{author}{\bibfnamefont{A.}~\bibnamefont{Boh{\'e}}},
  \bibinfo{author}{\bibfnamefont{L.}~\bibnamefont{Haegel}},
  \bibinfo{author}{\bibfnamefont{S.}~\bibnamefont{Husa}}, \bibnamefont{et~al.},
  \bibinfo{journal}{Phys.Rev.Lett.} \textbf{\bibinfo{volume}{113}},
  \bibinfo{pages}{151101} (\bibinfo{year}{2014}), \eprint{1308.3271}.

\bibitem[{\citenamefont{Sathyaprakash and
  Dhurandhar}(1991)}]{Sathyaprakash:1991mt}
\bibinfo{author}{\bibfnamefont{B.~S.} \bibnamefont{Sathyaprakash}}
  \bibnamefont{and} \bibinfo{author}{\bibfnamefont{S.~V.}
  \bibnamefont{Dhurandhar}}, \bibinfo{journal}{Phys. Rev. D}
  \textbf{\bibinfo{volume}{44}}, \bibinfo{pages}{3819} (\bibinfo{year}{1991}).

\bibitem[{\citenamefont{Owen}(1996)}]{Owen:1995tm}
\bibinfo{author}{\bibfnamefont{B.~J.} \bibnamefont{Owen}},
  \bibinfo{journal}{Phys. Rev. D} \textbf{\bibinfo{volume}{53}},
  \bibinfo{pages}{6749} (\bibinfo{year}{1996}), \eprint{gr-qc/9511032}.

\bibitem[{\citenamefont{Owen and Sathyaprakash}(1999)}]{Owen:1998dk}
\bibinfo{author}{\bibfnamefont{B.~J.} \bibnamefont{Owen}} \bibnamefont{and}
  \bibinfo{author}{\bibfnamefont{B.}~\bibnamefont{Sathyaprakash}},
  \bibinfo{journal}{Phys.Rev.} \textbf{\bibinfo{volume}{D60}},
  \bibinfo{pages}{022002} (\bibinfo{year}{1999}), \eprint{gr-qc/9808076}.

\bibitem[{\citenamefont{Cokelaer}(2007)}]{hexabank}
\bibinfo{author}{\bibfnamefont{T.}~\bibnamefont{Cokelaer}},
  \bibinfo{journal}{Phys.~Rev.~D} \textbf{\bibinfo{volume}{76}},
  \bibinfo{pages}{102004} (\bibinfo{year}{2007}), \eprint{arXiv:0706.4437}.

\bibitem[{\citenamefont{Babak et~al.}(2006)\citenamefont{Babak,
  Balasubramanian, Churches, Cokelaer, and Sathyaprakash}}]{Babak:2006ty}
\bibinfo{author}{\bibfnamefont{S.}~\bibnamefont{Babak}},
  \bibinfo{author}{\bibfnamefont{R.}~\bibnamefont{Balasubramanian}},
  \bibinfo{author}{\bibfnamefont{D.}~\bibnamefont{Churches}},
  \bibinfo{author}{\bibfnamefont{T.}~\bibnamefont{Cokelaer}}, \bibnamefont{and}
  \bibinfo{author}{\bibfnamefont{B.}~\bibnamefont{Sathyaprakash}},
  \bibinfo{journal}{Class.Quant.Grav.} \textbf{\bibinfo{volume}{23}},
  \bibinfo{pages}{5477} (\bibinfo{year}{2006}), \eprint{gr-qc/0604037}.

\bibitem[{\citenamefont{Brown et~al.}(2012)\citenamefont{Brown, Harry,
  Lundgren, and Nitz}}]{Brown:2012qf}
\bibinfo{author}{\bibfnamefont{D.~A.} \bibnamefont{Brown}},
  \bibinfo{author}{\bibfnamefont{I.}~\bibnamefont{Harry}},
  \bibinfo{author}{\bibfnamefont{A.}~\bibnamefont{Lundgren}}, \bibnamefont{and}
  \bibinfo{author}{\bibfnamefont{A.~H.} \bibnamefont{Nitz}}
  (\bibinfo{year}{2012}), \eprint{1207.6406}.

\bibitem[{\citenamefont{Harry et~al.}(2013)\citenamefont{Harry, Nitz, Brown,
  Lundgren, Ochsner et~al.}}]{Harry:2013tca}
\bibinfo{author}{\bibfnamefont{I.}~\bibnamefont{Harry}},
  \bibinfo{author}{\bibfnamefont{A.}~\bibnamefont{Nitz}},
  \bibinfo{author}{\bibfnamefont{D.~A.} \bibnamefont{Brown}},
  \bibinfo{author}{\bibfnamefont{A.}~\bibnamefont{Lundgren}},
  \bibinfo{author}{\bibfnamefont{E.}~\bibnamefont{Ochsner}},
  \bibnamefont{et~al.} (\bibinfo{year}{2013}), \eprint{1307.3562}.

\bibitem[{\citenamefont{Babak}(2008)}]{Babak:2008rb}
\bibinfo{author}{\bibfnamefont{S.}~\bibnamefont{Babak}},
  \bibinfo{journal}{Class.Quant.Grav.} \textbf{\bibinfo{volume}{25}},
  \bibinfo{pages}{195011} (\bibinfo{year}{2008}), \eprint{0801.4070}.

\bibitem[{\citenamefont{Harry et~al.}(2009)\citenamefont{Harry, Allen, and
  Sathyaprakash}}]{Harry:2009}
\bibinfo{author}{\bibfnamefont{I.~W.} \bibnamefont{Harry}},
  \bibinfo{author}{\bibfnamefont{B.}~\bibnamefont{Allen}}, \bibnamefont{and}
  \bibinfo{author}{\bibfnamefont{B.}~\bibnamefont{Sathyaprakash}},
  \bibinfo{journal}{Phys.Rev.} \textbf{\bibinfo{volume}{D80}},
  \bibinfo{pages}{104014} (\bibinfo{year}{2009}), \eprint{0908.2090}.

\bibitem[{\citenamefont{{Manca} and {Vallisneri}}(2010)}]{2010PhRvD81b4004M}
\bibinfo{author}{\bibfnamefont{G.~M.} \bibnamefont{{Manca}}} \bibnamefont{and}
  \bibinfo{author}{\bibfnamefont{M.}~\bibnamefont{{Vallisneri}}},
  \bibinfo{journal}{\prd} \textbf{\bibinfo{volume}{81}}, \bibinfo{eid}{024004}
  (\bibinfo{year}{2010}), \eprint{0909.0563}.

\bibitem[{\citenamefont{Ajith et~al.}(2014)\citenamefont{Ajith, Fotopoulos,
  Privitera, Neunzert, and Weinstein}}]{Ajith:2012mn}
\bibinfo{author}{\bibfnamefont{P.}~\bibnamefont{Ajith}},
  \bibinfo{author}{\bibfnamefont{N.}~\bibnamefont{Fotopoulos}},
  \bibinfo{author}{\bibfnamefont{S.}~\bibnamefont{Privitera}},
  \bibinfo{author}{\bibfnamefont{A.}~\bibnamefont{Neunzert}}, \bibnamefont{and}
  \bibinfo{author}{\bibfnamefont{A.}~\bibnamefont{Weinstein}},
  \bibinfo{journal}{Phys.Rev.} \textbf{\bibinfo{volume}{D89}},
  \bibinfo{pages}{084041} (\bibinfo{year}{2014}), \eprint{1210.6666}.

\bibitem[{\citenamefont{Robinson et~al.}(2008)\citenamefont{Robinson,
  Sathyaprakash, and Sengupta}}]{Robinson:2008un}
\bibinfo{author}{\bibfnamefont{C.~A.~K.} \bibnamefont{Robinson}},
  \bibinfo{author}{\bibfnamefont{B.~S.} \bibnamefont{Sathyaprakash}},
  \bibnamefont{and} \bibinfo{author}{\bibfnamefont{A.~S.}
  \bibnamefont{Sengupta}}, \bibinfo{journal}{Phys. Rev.}
  \textbf{\bibinfo{volume}{D78}}, \bibinfo{pages}{062002}
  (\bibinfo{year}{2008}), \eprint{0804.4816}.

\bibitem[{\citenamefont{Tanaka and Tagoshi}(2000)}]{Tanaka:2000xy}
\bibinfo{author}{\bibfnamefont{T.}~\bibnamefont{Tanaka}} \bibnamefont{and}
  \bibinfo{author}{\bibfnamefont{H.}~\bibnamefont{Tagoshi}},
  \bibinfo{journal}{Phys.Rev.} \textbf{\bibinfo{volume}{D62}},
  \bibinfo{pages}{082001} (\bibinfo{year}{2000}), \eprint{gr-qc/0001090}.

\bibitem[{\citenamefont{Keppel et~al.}(2013)\citenamefont{Keppel, Lundgren,
  Owen, and Zhu}}]{Keppel:2013kia}
\bibinfo{author}{\bibfnamefont{D.}~\bibnamefont{Keppel}},
  \bibinfo{author}{\bibfnamefont{A.~P.} \bibnamefont{Lundgren}},
  \bibinfo{author}{\bibfnamefont{B.~J.} \bibnamefont{Owen}}, \bibnamefont{and}
  \bibinfo{author}{\bibfnamefont{H.}~\bibnamefont{Zhu}},
  \bibinfo{journal}{Phys.Rev.} \textbf{\bibinfo{volume}{D88}},
  \bibinfo{pages}{063002} (\bibinfo{year}{2013}), \eprint{1305.5381}.

\bibitem[{\citenamefont{Mohanty}(1998)}]{PhysRevD.57.630}
\bibinfo{author}{\bibfnamefont{S.~D.} \bibnamefont{Mohanty}},
  \bibinfo{journal}{Phys. Rev. D} \textbf{\bibinfo{volume}{57}},
  \bibinfo{pages}{630} (\bibinfo{year}{1998}),
  \urlprefix\url{http://link.aps.org/doi/10.1103/PhysRevD.57.630}.

\bibitem[{LAL()}]{LAL}
\emph{\bibinfo{title}{{LSC} {A}lgorithms {L}ibrary}},
  \urlprefix\url{https://www.lsc-group.phys.uwm.edu/daswg/projects/lalsuite.html}.

\bibitem[{aLI(2010)}]{aLIGOPSD}
\emph{\bibinfo{title}{Advanced ligo anticipated sensitivity curves}}
  (\bibinfo{year}{2010}),
  \urlprefix\url{https://dcc.ligo.org/LIGO-T0900288/public}.

\bibitem[{\citenamefont{Gaur et~al.}()\citenamefont{Gaur, Sengupta
  et~al.}}]{Gaur:2015aa}
\bibinfo{author}{\bibfnamefont{G.}~\bibnamefont{Gaur}},
  \bibinfo{author}{\bibfnamefont{A.}~\bibnamefont{Sengupta}},
  \bibnamefont{et~al.}, \bibinfo{note}{in preparation}.

\bibitem[{\citenamefont{Privitera et~al.}(2013)\citenamefont{Privitera,
  Mohapatra, Ajith, Cannon, Fotopoulos et~al.}}]{Privitera:2013xza}
\bibinfo{author}{\bibfnamefont{S.}~\bibnamefont{Privitera}},
  \bibinfo{author}{\bibfnamefont{S.~R.~P.} \bibnamefont{Mohapatra}},
  \bibinfo{author}{\bibfnamefont{P.}~\bibnamefont{Ajith}},
  \bibinfo{author}{\bibfnamefont{K.}~\bibnamefont{Cannon}},
  \bibinfo{author}{\bibfnamefont{N.}~\bibnamefont{Fotopoulos}},
  \bibnamefont{et~al.} (\bibinfo{year}{2013}), \eprint{1310.5633}.

\end{thebibliography}

\end{document}